# A Business Goal Driven Approach for Understanding and Specifying Information Security Requirements

Xiaomeng Su[1], Damiano Bolzoni[2], Pascal van Eck[1], and Roel Wieringa[1]

[1] University of Twente, Information Systems Group, Enschede, The Netherlands
x.su@ewi.utwente.nl vaneck@ewi.utwente.nl r.j.wieringa@ewi.utwente.nl
[2] University of Twente, Distributed and Embedded System Group, Enschede, The Netherlands
damiano.bolzoni@utwente.nl

**Abstract.** We present a new approach to explicity link security requirements with the organization's business vision. A conceptual framework is presented, where the relationships between business vision, critical impact factors and valuable assets (together with their security requirements) are clearly show. Evaluating this relationships and considering also organization's business drivers, it is possible to define a plan for prioritizing security requirements primary required by organization's business.

## 1 Introduction

Security has long been a consideration for those responsible for computing systems. Over the years, the level of threat and the need for corresponding safeguards have varied, depending on the environment in which the systems were deployed and the data they processed. However, the Internet marked the beginning of a new social and commercial phenomenon. It also opened a door for those seeking to electronically exploit others' data and infrastructure for their own purposes. A recent CSI/FBI computer crime and security survey has showed that 56% of the respondents experienced unauthorized use of their computer systems in the past 12 months. The same survey also reported that the total losses for 2005 were $ 130,104,542 for the 639 respondents that were willing and able to estimate losses, in other words, $203.606 losses per respondent. The past decade has also seen the introduction of legislation that often places a responsibility on organizations to demonstrate a requisite level of care when handling computer data.

The increasing concerns of clients, particularly in online commerce, plus the impact of legislations on information security have compelled companies to put more resources in information security. It is clear that senior managers in many organizations are now expressing a much greater interest in information security. Understanding and specifying what kind of security an organization need is however a difficult task. Many underlying goals (why and what security is needed) remain tacit within organizations requirements and end up being articulated as specifications of the security control baseline (how security will be achieved) without a clear rationale. The problem becomes more urgent when more and more organizations are involved in collaboration and commerce. Being able to articulate security goals and requirement consistently, based on an accurate view of existing security capabilities, and using shared understandings, becomes much more important. Networked business will be difficult to function if the organizations involved cannot agree: why security is necessary; the scope it should cover and what each organization expects it to achieve.

The complexity of undertaking an enterprise wide view of security management can be illustrated in the challenges facing chief security officers (CSO). Often CSOs are tasked with "securing" the organization, but may not be clear on what that means. As a result, the CSO is often left to answers of very important organization questions without specific guidance:

- What needs to be secured? Why, and in what priority?
- How to ensure that people agree on the above issue?
- How will I know when the organization has been "secured"? What will be used to measure success?

We believe to answer the above questions, it is necessary to link the security requirements with the organization's unique business drivers. For a production company, the availability of its production control system is of vital importance, whereas for a financial service provider, it is important to protect the integrity of its financial transactions. Different organizations have different business drivers, which in turn determined their different requirements to security. This should be taken into consideration while understanding and specifying security requirements. It is therefore important to develop techniques and instruments to help stakeholders articulating the connection between security requirements and the business drivers in a systematic way.

## 2 Formulating and Understanding Security Goals and Requirements

A security requirement specification tells what should be secured and why. It identifies the organizations' needs with respect to security. Considere, for example, the differences between the needs of a university and those of a cryptographic organization. The university fosters scholarship and open research: papers, discoveries, and work are available to the general public as well as to other academics. The cryptographic organization, on the other hand, prizes secrecy. The university will need to protect the integrity and confidentiality of the data, such as grades, on its systems. It might also want to ensure that the system is available via the Internet so that students, faculty, and other researchers have access to information. The cryptographic organization, though, will emphasize confidentiality of all its work.

When an organization wants to secure its system, it must first determine what requirements to meet. Given that organizations normally have limited resources to protect its assets, it is equally important to determine which requirements are more important and thus should be prioritized. To achieve this, we propose to use a conceptual framework where security requirements are linked to the unique business drivers of the organization in question. Figure 1 portrays the conceptual framework. Business vision are high level business goals the organization has. Critical Impact Factors (CIFs) identify what will be the business impacts if security requirements are violated. Valuable assets and their security requirements are inventories of security requirements. Valuable assets and their security requirements have an effect on the CIFs and the CIFs in turn impact the accomplishment of the organization's business vision. In other words, we can use organization's business vision to prioritize the CIFs, which further can be used to prioritize the security requirements. To achieve that, three subsequent steps need to be taken. Firstly we need to enumerate valuable assets and their security requirements. Secondly, an organization's CIFs and business vision need to be defined. Thirdly, security requirements shall be linked with CIFs and business vision. We will discuss them in detail.

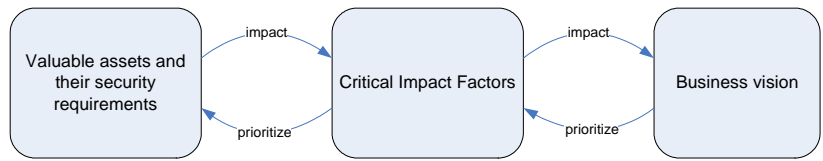

**Fig. 1.** Linking security requirements with business vision via CIF.

### 2.1 The business vision

Each organization has its own unique business vision that defines the very principles of how the business wants to achieve its goals. This vision, moreover, often changes over time to reflect changing circumstances. Notwithstanding this diversity, scholars in business administration have identified certain "paterns" in the business vision of leading firms. In this paper, we use the well-known *value disciplines* identified by Treacy and Wiersema [8,9] as a frameworks for understanding the business vision.

Treacy and Wiersema argue that there are three generic ways a business can differentiate itself from the competition, which they call *operational excellence*, *customer intimacy*, and *product leadership*. Each of these three value disciplines aims at creating distinguishing value for customers in a different way. A company striving for operational excellence focuses on offering its products with the least possible hassle (usually, at the lowest cost) to its customers. A customer intimacy company aims at delivering exactly what its customers want by investigating the needs of a narrow market and customizing its offerings to this market. Finally, a product leader aims at delivering radically innovative products that create an unbridgable gap with the competition.

Each of the three value disciplines leads to a radically different operating model for the company: the culture, processes, management systems and IT systems of the company. For instance, while operational excellence calls for highly standardized business processes, the customer intimacy discipline requires just the opposite: to meet customer requirements, business processes should be as flexible as possible. Security requirements should be likewise aligned with the requirements imposed upon culture, processes and management systems by the value discipline choosen.

### 2.2 Identifying the critical impact factors

When security incidents happen, they may lead to damage to organizations. Critical impact factors (CIFs) are the indicators about what kind of damage the security incidents incur to the organization. They can include those within the control of the organization, as well as some that the organization may not be able to fully control (e.g., legal liability, finance losses and reputational damage).

The identification activity of CIFs is also referred as business impact analysis (BIA): it is important to remark the difference between BIA and risk analysis.

BIA identifies the critical business functions within the organization and determines the effect of external and internal CIFs upon the various parts of organization, using a combination of relevant literature study and specific case studies in the organizations.

Risk analysis involves identifying the most probable threats to an organization and analyzing the related vulnerabilities of the organization to those threats.

Figure 2 illustrate an example list of critical impact factors.

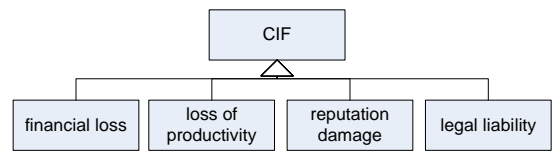

**Fig. 2.** An example of Critical Impact Factors.

### 2.3 Selecting valuable assets and security requirements

The business vision can be used to guide the selection of valuable assets. Surely, the assets that are critical for accomplishing the business vision are the valuable ones for the organizations. For example, a financial service company that focuses on customer intimacy will consider its customer relationship management (CRM) systems as extremely valuable, while a financial service company that focuses at product leadership will likely value its systems for developing new financial products even higher.

Information security has been defined as encompassing systems and procedures designed to ensure the confidentiality, integrity and availability [3] of an organization's critical information and technical assets. Information assets represent the data and information, in either physical or electronic from, that is critical to the organization. Technical assets represent those assets that support the storage, transmission, and processing of data and information and therefore are important to transforming data and information to be used by the organization. People can be an asset to the organization as well for similar reason – they can be a primary way of storing, transporting, or processing data.

So, IT security is about safeguard certain desired properties. The core of computer and information security is widely regarded as the preservation of three factors: confidentiality (ensuring that information is accessible only to those authorized to access), integrity (safeguarding the accuracy and completeness of information and processing methods) and availability (ensuring that authorized users have access to information and associated assets when required) [3].

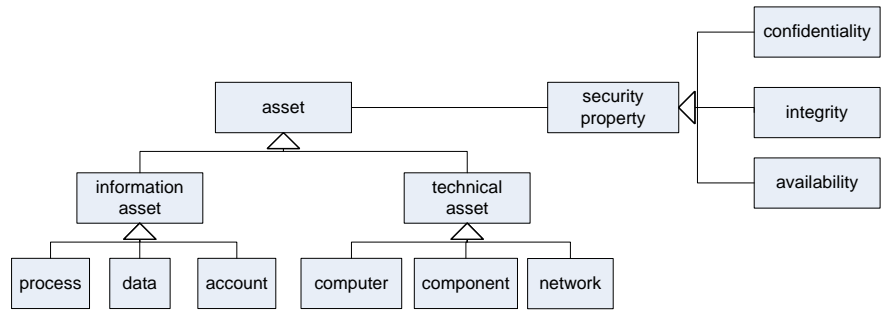

**Fig. 3.** A simple ontology of asset and security property.

Figure 3 depicts a simple ontology of asset and the security properties that are in the scope. Such an ontology can be used as a starting point to structure assets and their security properties. It is a minimum set and can be extended. For instance, some will include privacy issues like anonymity as a security property too. Using such an ontology, the assets and their security properties can be structured accordingly. Figure 4 gives an example.

---

[3] some authorities treat communication security issues such as non-repudiation and privacy related issues such as anonymity as additional aspects of security

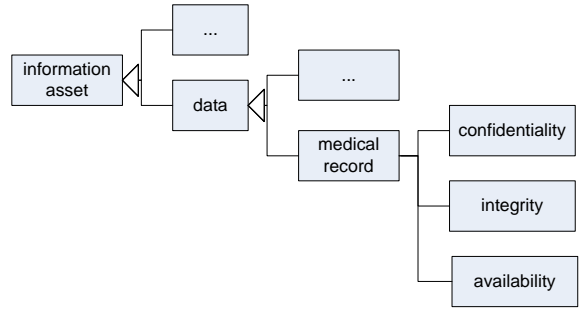

**Fig. 4.** An example of assets and their security properties.

### 2.4 Prioritizing security requirements

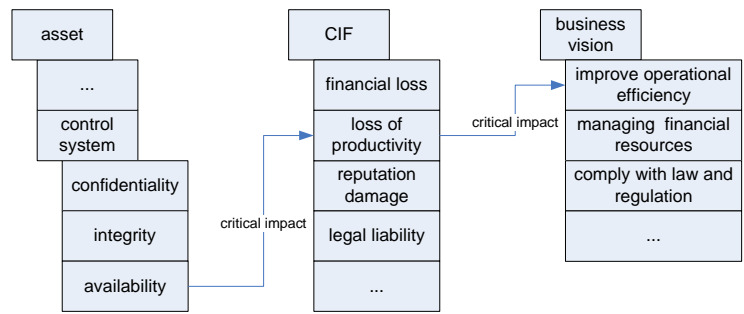

**Fig. 5.** An example of linking asset security requirement with business vision via CIFs.

To further elaborate the relations between security requirements and the business vision, the connection between them can be established via the linkage of CIFs. Figure 5 provides an example of such linkage for a production company. In this example, the organization is stating that any compromise to "availability" of the "control system" has "critical impact" to "loss of productivity", which in turn has "critical impact" to the organization's vision "improve operational efficiency". The impact severity can be categorized according to the organization's needs. An example categorization can be *critical impact*, *marginal impact*, and *negligible impact*. In this way, each security requirements can be connect its CIFs and the CIFs further to business vision.

Using the impact diagram like figure 5, it is possible to categorize and prioritize the different security requirements. Requirements that have "critical impacts" on CIFs, which have "critical impacts" on business vision, should be considered of most importance. These requirements shall be satisfied first if the resources (time, money, manpower etc.) are limited. In this example, it means for the production company, it is more important to mitigate threats to the control system's availability than for instance, threats to the control system's confidentiality. It is possible that one requirement may be linked to more than one CIFs. When that happens the overall significance of that security requirement can be determined by a number of ways. For example, one can choose the maximum impact level, e.g. if control system's availability not only "critical impacts" on loss of productivity but also has "marginal impact" on reputation damage, the overall impact should be "critical impacts". Alternatively, one can choose the average impact level. The organization shall decide which combination methods best reflects its situation.

The reason why we use CIFs to link critical assets and their security requirements with business goals is twofold. Business goals typically reside at strategic level. When the business goals are outlined, the stakeholders do not normally have a security focus in mind. The Critical Impact Factors on the other hand, reflect the business implication when security is compromised. It is of course possible to directly connect assets' security requirements with business goals. But then the shift of focus from pure technical level security concern to strategical level business concern seems abrupt. The introduction of CIFs makes the shift smooth and the line of reasoning easier to follow.

Once the requirements are categorized and prioritized. Other techniques, like attacks trees or misuse cases can be used to explore all possible threats and attack paths that would lead to the violation of security properties. In this example, it is to find out how the control system's availability can be compromised. Our approach is complementary to this line of work, in the sense that we provides a business grounded rational for why certain security requirements are important while others are not.

## 3   Related Research

There exists a number of security standards, among which COBIT (Control Objectives for Information and related Technology) [2] and BS7799 [1] are of particular relevance to our work. COBIT is an international de-facto standard for information control and IT risk management, addressing IT governance and control practices. It provides a reference business-oriented framework for management, users and control and security auditors. COBIT defines control objectives but does not provide guidelines on how to reach the objectives. BS7799 (originally published in 1995) is strictly focused on IT security and it is divided into two parts. The first part *Code of Practice for Information Security Management* became an official ISO standard, ISO 17799, in 2000. It contains general security guidelines including policies, practices, procedures, organizational structures and software functions. The second part *Information Security Management - Specifications* became a ISO standard, ISO 27001, in 2005. It contains technical requirements. ISO17799/27001 addresses company's security from a best practice point of view, which does not provide any answer to why certain security mechanisms are in place for a particular organization. Both COBIT and ISO17799 however, do not define guidelines on how to prioritize in a proper way the company assets and their security properties.

Our approach is also related to the work of security requirement modeling. Sinder and Opdahl [7] suggested to have negative use cases or scenarios in connection with security. Yu and Liu [10] in their work shows actors and misactors having contradictory goals in an extension of $i^*$ diagrams [4]. Links such as "break" or "hurt" can show how an attack prevents the legitimate users from reaching their goals, or how countermeasures can thwart an attacks. This line of work focuses on how to model threat, including the threat actors and their attack paths. Our approach on the other hand, focuses on providing business rationale for explaining why certain security requirements exist in the first place as well as explaining why some security requirements are more important than others. Our approach can be combined with the modeling work. First, the security requirements are ranked using our approach. Next, for the prioritized security requirements, misuse case or attack trees [6] can be use to model how attacks that will violate the security requirements could actually happen.

Finally, our work is also related to the work of security risk assessment. Traditionally, in risk assessment methodologies, e.g. OCTAVE [5] (Operationally Critical Threat, Asset and Vulnerability Evaluation), risk of each threat is determined by the impact of the threat once it happens and how likely it will happen [REF]. For the impact, a severity level (e.g. high, low or medium) is assigned. What often happens is that the severity level is assigned based on the person's own experience without explicit rationale. Our approach can enhance this process because it provides an explicit tie to the organization's business drivers.

## 4 Conclusions

In this paper, we argued the necessity of making explicit the tie between security requirements and the organization's business drivers. A company like Google that is making profit with online advertising would tend to keep its search engine availability as high as possible, without taking too much care about data confidentiality (thus is, they are public available with few clicks). On the other hand, a financial entity that offers online servicing would be more focused on data confidentiality than availability because possible legal acts brought by damaged users could cost more than a temporary service shutdown. The difference is determined by the underlying different business drivers the different organizations have.

Further we proposed a conceptual framework to that aim. Three main elements are included in the framework, namely, business vision, CIFs and valuable assets and their security requirements. The connection between business goals and security requirements, once established, can be used to provide rational for prioritizing security requirements. A number of issues need future work. We need to define methods that help to come up with a proper set of CIFs. Also in the face of contradicting business visions, proper guidelines should be given on how to combine the results.

## References

1. BS 7799/ISO 17799 Information Security, 2000. URL http://www.bsi-global.com/Global/bs7799.xalter.
2. CobiT: Control Objectives for Information and related Technology. URL http://www.isaca.org.
3. S. Furnell. *Computer Insecurity – risking the system*. Springer, 2005.
4. J. Mylopoulos, L. Chung, and E. Yu. From object-oriented to goal-oriented requirement analysis. *Communications of the ACM*, 42(1):31–37, January 1999.
5. OCTAVE risk methodology. URL http://www.cert.org/octave/.
6. B. Schneier. *Secrets and Lies: Digital Security in a Networked World*. John Wiley & Sons, 2000.
7. G. Sindre and A. L. Opdahl. Eliciting security requirements with misuse cases. *Requirement Engineering*, 10(1):34–44, 2005.
8. M. E. Treacy and F. D. Wiersema. Customer Intimacy and Other Value Disciplines. *Harvard Business Review*, 71(1):84–93, jan 1993.
9. M. E. Treacy and F. D. Wiersema. *The Discipline of Market Leaders: Choose Your Customers, Narrow Your Focus, Dominate Your Market*. Perseus Publishing, 1997.
10. E. Yu and L. Liu. Modelling trust in $i^*$ strategic actors framework. In *Proceedings of the third workshop on deception, fraud and trust in agent societies*, June 2000.